# Proposing a System Level Machine Learning Hybrid Architecture and Approach for a Comprehensive Autism Spectrum Disorder Diagnosis

Ryan Liu[1] and Spencer He[2]

**Abstract** - Autism Spectrum Disorder (ASD) is a severe neuropsychiatric disorder that affects intellectual development, social behavior, and facial features, and the number of cases is still significantly increasing. Due to the variety of symptoms ASD displays, the diagnosis process remains challenging, with numerous misdiagnoses as well as lengthy and expensive diagnoses. Fortunately, if ASD is diagnosed and treated early, then the patient will have a much higher chance of developing normally. For an ASD diagnosis, machine learning algorithms can analyze both social behavior and facial features accurately and efficiently, providing an ASD diagnosis in a drastically shorter amount of time than through current clinical diagnosis processes. Therefore, we propose to develop a hybrid architecture fully utilizing both social behavior and facial feature data to improve the accuracy of diagnosing ASD. We first developed a Linear Support Vector Machine for the social behavior based module, which analyzes Autism Diagnostic Observation Schedule (ADOS) social behavior data. For the facial feature based module, a DenseNet model was utilized to analyze facial feature image data. Finally, we implemented our hybrid model by incorporating different features of the Support Vector Machine and the DenseNet into one model. Our results show that the highest accuracy of 87% for ASD diagnosis has been achieved by our proposed hybrid model. The pros and cons of each module will be discussed in this paper.

1. Introduction

Neuropsychiatric disorders account for about 14% of the global burden of disease, and Autism Spectrum Disorder (ASD) is a substantial part of that [1]. The number of ASD cases has risen tremendously, by 178%, since the year 2000, and it now affects 31 million people, including 1 in every 54 children in the United States alone [2,3]. Although there is no known cure for ASD, if it is recognized early, treatment can tremendously help the patient. Unfortunately, the diagnosis process for ASD can be quite time-consuming and expensive, lasting almost 3 years and costing around $60,000 per year per patient for a complete diagnosis, so it is very rare to see early intervention. These hindrances will only further prevent patients from receiving an accurate ASD diagnosis in a timely fashion. Many of these issues stem from the current clinical diagnosis processes. Like most behavioral and mental health conditions, ASD has a variety of symptoms that are diagnosed through behavioral examinations in four main steps, as shown in Figure 1. Figure 1 shows a diagram of current clinical procedures in an ASD diagnosis, consisting of four major steps:

developmental monitoring, physical screening, diagnostic checklists, and other supplemental tests. Developmental monitoring is made up of observations during periodic health checkups. Specialists then administer routine physical tests, including checking the patient's eyes, ears, elementary motor skills, etc, in order to make sure there are no physical issues. For example, if a patient does not respond to a parent's call, doctors need to make sure the patient is not deaf instead of assuming that the patient has ASD. Then, a patient's behavior will be assessed through the use of checklists, like Autism Diagnostic Observation Schedule, Autism Diagnostic Interview-Revised, Modified Checklist for Autism in Toddlers, and more. Finally, specialists complete other supplemental tests with an in-person evaluation [4-6].

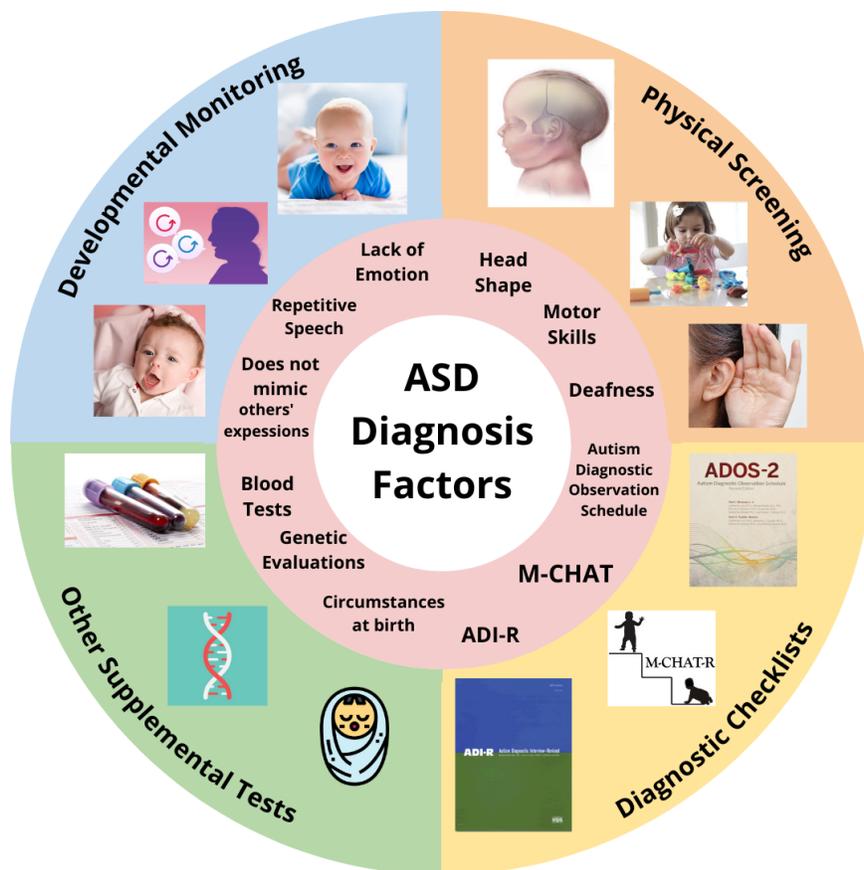

Figure 1: Diagram of current clinical procedures in an ASD diagnosis, consisting of four major steps: 1. Developmental Monitoring; 2. Physical Screening; 3. Diagnostic Checklists; and 4. Other Supplemental Tests. These 4 steps analyze both the social behavior and the facial features of the patient, and there are occasional supplemental tests as well to look into other possible factors that may contribute to the disorder.

Table 1: Compares basic information, including models used, main results, and limitations, of the past works reviewed in this research.

| Research Paper Title | Main Purpose | Model Used | Datasets | Main Results | Limitations |
|---|---|---|---|---|---|
| **Use of machine learning to shorten observation-based screening and diagnosis of autism[7]** | Analysis of the ADOS - AGRE dataset | 16 different classifiers(ADT, Ridor, FilteredClassifier, other decision trees) | ADOS AGRE dataset AC Simons | The ADTree was the most effective in terms of sensitivity, specificity, and accuracy. | Model accuracies are not replicable, Limited quality of data |
| **Use of Artificial Intelligence to Shorten the Behavioral Diagnosis of Autism[8]** | Analysis of the ADI-R dataset | 15 different classifiers(ADT, Ridor, FilteredClassifier, other decision trees) | ADI-R dataset 895 - Autism 75 - Without SSC AC | The ADTree classifier with 93% fewer questions and 99% accuracy was effective in speeding up Autism diagnosis | Data is unbalanced which may cause a bias in the model |
| **Applying Machine Learning to Facilitate Autism Diagnosis Pitfalls and promises[9]** | To debunk the findings of the previous two papers | 16 different classifiers(ADT, Ridor, FilteredClassifier, other decision trees) | ADI-R dataset and ADOS dataset | Both papers included inaccuracies in problem formation and interpretations | Did not discuss machine learning models |
| **Mobile detection of autism through ML on home video: A development and prospective validation study[10]** | To determine whether mobile autism diagnosis is effective and which model is best | ADTree8, ADTree7, SVM5, LR5, LR9, SVM12, SVM10, LR10 | ADOS + 193 home videos | Autism can be diagnosed with a >90% accuracy on mobile platforms | ADOS is not usually used alone to determine a diagnosis |
| **Machine Learning Classifiers for Autism Spectrum Disorder: A Review[11]** | Reviewing the application of machine learning in Autism diagnosis | SVMs, Decision Trees, Naive Bayes, Random forests, Logistic Regression, K-Nearest Neighbors | ADDM, ADI-R, SRS, ADOS | The most widely used algorithm is SVM to accelerate and improve diagnosis accuracy of ASD | Multiple modules used which may introduce interference for accuracy |

Similar to the new diagnosis process of other diseases, ASD has seen many diagnosis attempts utilizing machine learning techniques in order to improve the inefficient diagnosis time of ASD. Researchers have relied on various machine learning models to diagnose ASD, such as Support Vector Machines, Decision Trees, Random Forests, and Logistic Regression models, as shown in Table 1. Although many of these models have proven successful in terms of model performance, like over 90% in accuracy or sensitivity and specificity, they still face numerous limitations in the application side of their work. For example, these machine learning solutions only analyze one form of data during their process--only social behavior data or facial feature data. This is a major issue because ASD affects multiple areas of the body, so only analyzing one area leaves out a significant portion of variability for medical specialists. Another limitation is the quantity and balance of data used to train and test their models. Data from the ADI-R database, for example, is very often used by researchers. However, it contains data from primarily ASD patients, with very little data from non-ASD patients. Therefore, this skew in the data may have potentially skewed the results of certain researchers. They also do not have a large quantity of data for training and evaluation of the models, which may skew the results. In this research, a hybrid model architecture has been proposed to utilize machine learning in the diagnosis of ASD. It will analyze all areas that ASD affects and incorporate all the information a doctor needs to make an informed ASD diagnosis [7-11].

2. Databases

This research utilizes three different databases: an Autism Diagnostic Observation Schedule (ADOS) database which provides social behavior data, a Kaggle database which provides facial feature image data, and our self-gathered home video database [13]. The ADOS and Kaggle data were used in order to incorporate all the information necessary for an ASD diagnosis, while the self-gathered home video database adds a unique and different source of data. First, our social behavior module analyzes the ADOS data. The entire system is focused on ADOS-2, which primarily diagnoses ASD among toddlers; therefore, the scope of our data is limited to preschool age children, with a mean age under 5. The ADOS evaluation essentially consists of certain tasks of varying complexity for the patient to accomplish, such as using blocks or telling a story. A specialist or doctor will then analyze certain factors that are typically indicative of ASD, like maladjusted eye contact or echolalia. These factors are then scored on a scale of 0 to 3 and 7 to 9, which measures the severity for the certain factor. For example, a score of 3 indicates that the behavior of type specified is present to a degree that interferes with functioning, but a score of 8 means that the behavior is unknown or missing [14]. In our system, we analyzed 5 of the most important features for an ASD diagnosis: echolalia, conversation skills, amount of maladjusted eye contact, facial expressions, and social responses. We used the social behavior information from 1319 different patients.

The facial feature based module analyzes data from the Kaggle database, which consists of facial images of ASD and non-ASD patients. The data is also limited to preschool age children, just like the data from our social behavior based module. The data is split into 90% for training and 10% for testing: 1268 images for training, 150 images for testing, and 100 images for validation, in both the ASD and non-ASD categories. Additionally, the images are all the same dimensions, 224 x 224 x 3 pixels, and are centered faces of the children [13].

In order to develop and test the hybrid model, we needed a paired database of ADOS social behavior and facial feature image data. Unfortunately, we were not able to find any publicly available paired databases that matched our criteria, even after extended research. For example, research from first author Q. Tariq has reported a video paired database, but it is not publicly available. Therefore, we culled the data ourselves by gathering 150 publicly available home videos from credible and verifiable resources, like the YouTube channel entitled "Autism family" [14]. This channel provides educational videos showcasing how ASD affects people daily. From these videos, we extracted the social behavior and facial feature information. We do not have professional expertise to score these videos based on the ADOS criteria, but Dr. Hui Qi Tong from the Stanford Department of Psychiatry and Behavioral Science was generous enough to analyze these videos and provide us with scores for the features we were analyzing. Then we were able to create our own database for our social behavior based module, the Numerical Social Behavior database (NSB). We then extracted about 300 images from these videos, which we input into another database for our facial feature based module, the Videos and Image Facial Feature database (VIFF).

## 3. Results and Discussion

In this research, the proposed hybrid architecture has been successfully developed, as shown in Figure 2. However, before we developed this architecture, we proposed two singular modules analyzing social behavior and facial features, separately. To fully understand the hybrid architecture, the first two modules must be addressed first

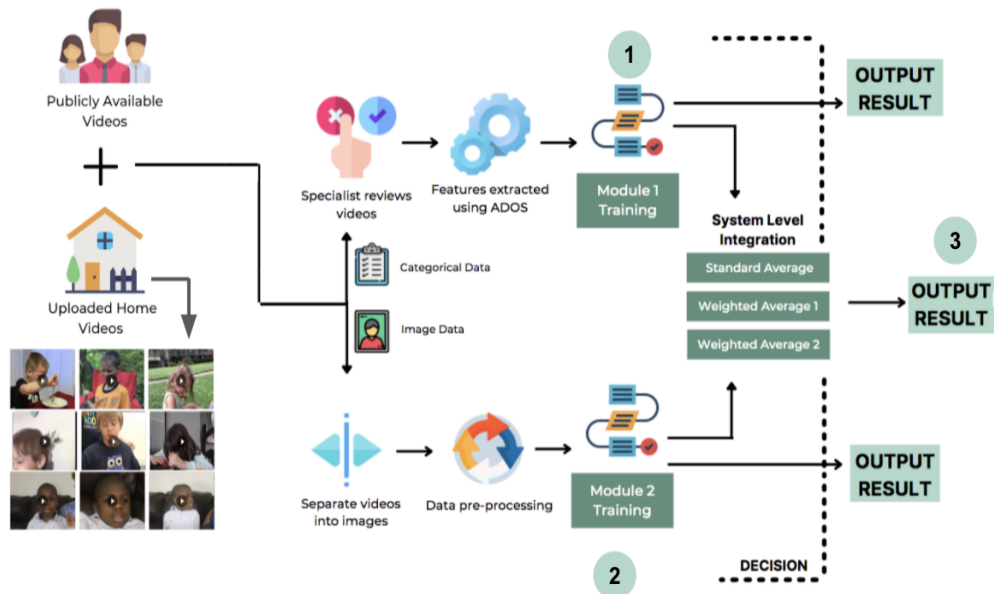

Figure 2: Displays the proposed ASD diagnosis method based on the ML hybrid architecture.

The Logistic Regression (LR) model's efficient runtime and capacity for experimentation allowed for us to design and develop it for our social behavior based module [15]. Our initial results were able to confirm those of previously highlighted works. However, we wanted to improve these results by tuning the learning rate, but the improvement was slight. Therefore, we developed a Linear Support Vector Machine (SVM) model because it is well suited for the size of our database. We began the training process by first splitting the ADOS data into 2 chunks: 80% of the data to be used for training and 20% of the data to be used for testing. After developing the Linear SVM model, our results further confirmed the results of previously highlighted works. After further optimizing the learning rate, the accuracy has improved by 2% and both sensitivity and precision has improved by 1%. Finding the perfect learning rate is important because it directly impacts the performance of the model. It has been found that if a learning rate is too low or too high, the accuracy will be low since the model will not be fitted correctly. Overall, the Linear SVM model we designed performed well, producing 96% accuracy, 96% sensitivity, and 77% precision, and it was able to surpass previously highlighted works, as shown in Table 2. The table showcases important metrics, including the accuracy, sensitivity, precision, medical source, dataset size, and the proportion of ASD to non-ASD patients, and it compares the proposed module with the previous works.

Table 2: Compares the results of the proposed social behavior module and previously highlighted works. It also highlights important metrics regarding the performance of the module.

| Statistic | Logistic Regression [10] | Logistic Regression [12] | Initial Logistic Regression | Post Optimization: Logistic Regression | Linear SVM[10] | Radial Kernel SVM[12] | Initial Linear SVM | Post Optimization: Linear SVM |
|---|---|---|---|---|---|---|---|---|
| Medical Record Source | 3 | 2,3 | 3 | 3 | 3 | 2,3 | 2 | 2 |
| Accuracy | 76% | 89% | 78% | 80% | 74% | 97% | 95% | 96% |
| Sensitivity | 100% | 98% | 77% | 78% | 95% | 97% | 95% | 96% |
| Precision | 76% | 98% | 72% | 73% | 71% | 97% | 77% | 77% |
| Dataset Size | 3143 | 4540 | 2870 | 2870 | 1319 | 4540 | 1319 | 1319 |
| $N_{ASD}/N_{non\text{-}ASD}$ | 2870/273 | 4189/273 | 2597/273 | 2597/273 | 1046/273 | 4189/343 | 1046/273 | 1046/273 |

For the facial feature based module, a Convolutional Neural Network (CNN) was initially developed because of its high efficiency and accuracy. Unfortunately, although the accuracy was very high, the model faced extreme overfitting issues, leading to poor performance in measuring the sensitivity and precision. Therefore, we developed a DenseNet model, which is a modified version of the CNN model. The data was split in the same format for the training of our facial feature based module: 80% for training, 20% for testing. Again, the initial results were able to confirm the results of other machine learning models. However, our optimization methods improved the model immensely. We first utilized a custom callback feature, which adjusted the learning rate and saved the best performing weights during training. Next, we extracted select images from the home videos we gathered and put those images into the validation dataset. Since the validation dataset is already used to tune and optimize the model, adding our own data would only improve the model [16]. Our optimization methods were very successful, outperforming the previous works in all aspects, including improving accuracy by 6%, sensitivity by 15%, and precision by 3%. Overall, the DenseNet model we designed produced a 92% accuracy, 91% sensitivity, and 89% precision, as shown in Table 3. The table highlights important metrics regarding the module's performance, including the accuracy, sensitivity, precision, dataset size, and the proportion of ASD to non-ASD patients, and it compares the proposed module with the previous works.

Table 3: Compares results of the proposed facial feature based module and previous works.

| Statistic | Oxford CNN | DNN | SVM | Random Forest | Random Forest[0] | Our Work: CNN | Initial DenseNet | Post Optimization: DenseNet |
|---|---|---|---|---|---|---|---|---|
| Age Range | 2-14 | 2-14 | 2-14 | 2-14 | 2-14 | 2-8 | 2-8 | 2-8 |
| Accuracy | 85% | 70% | 65% | 63% | 77% | 99% | 81% | 92% |
| Sensitivity | 76% | 74% | 68% | 69% | 82% | 66% | 81% | 91% |
| Precision | 87% | 63% | 62% | 58% | 73% | 66% | 83% | 89% |
| Dataset Size | 2940 | 3500 | 3500 | 3500 | 500 | 2836 | 3652 | 2836 |
| $N_{ASD}/N_{non-ASD}$ | 1470/1470 | 2641/859 | 2641/859 | 2641/859 | 250/250 | 1418/1418 | 1828/1824 | 1418/1418 |

After developing the social behavior and facial feature based modules, we designed our system-level hybrid model, which utilizes the performance of our social behavior and facial feature based modules to create a more comprehensive and equally accurate model. Since this model combines aspects from the two previous modules, we needed a paired dataset. This model used our paired database to test and produce the results because there were no publicly available paired datasets that were also balanced. We were able to extract the social behavior information from each patient in our database, as mentioned in the previous database section, and we extracted around 300 images (facial feature data) from these videos as well. In order to design the hybrid model, we first trained and tested our social behavior and facial feature based modules with our paired database. The results were lower, but that is likely due to the paired database not being as large compared to the ADOS or Kaggle databases. Then, we gathered the prediction vectors of each module, which report whether the patient has ASD or not, and these vectors are based on the patient data. For example, it is based on the number of images used for each patient. Finally, we combined these vectors to get our final output. A diagram of this architecture is shown in Figure 3, showcasing all the features and layers utilized in our hybrid model.

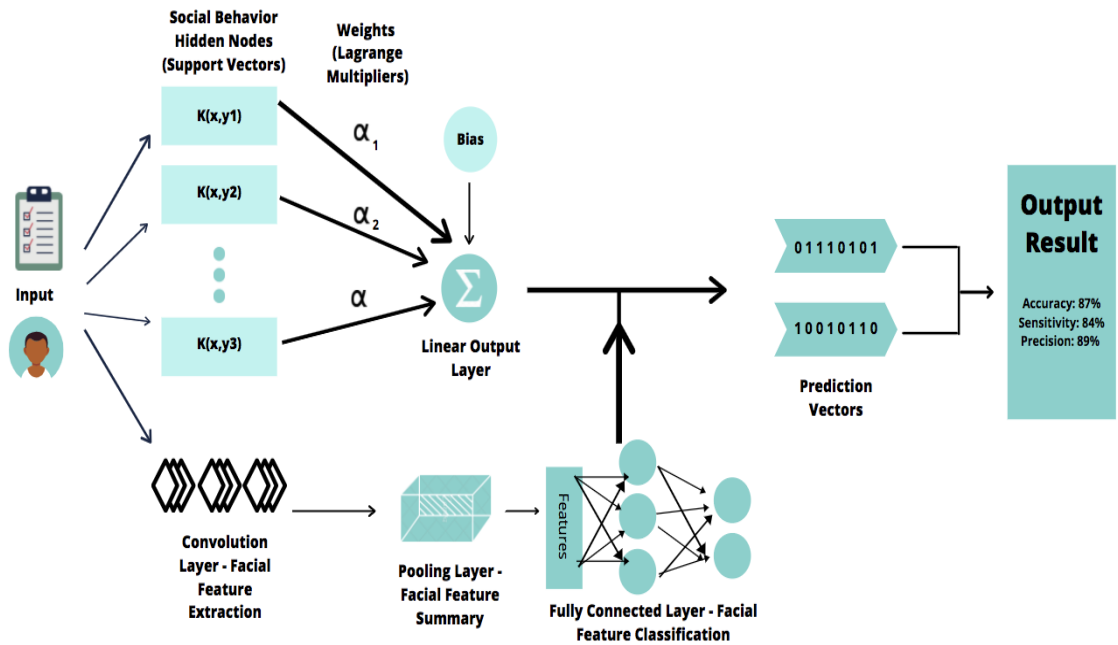

Figure 3: Displays the actual architecture of the proposed machine learning hybrid model for ASD diagnosis. Shows the layers and weights used in the coding of the proposed machine learning hybrid model.

After this test, we concluded that weighting our hybrid model based on the number of ASD patients in the training data provided the best results: 87% accuracy, 84% sensitivity, and 89% precision, as shown in Table 4. Table 4 is a summary of results from the three proposed modules in this work, the social behavior based module, the facial feature based module, and the hybrid model. The table compares the performance from these three modules, showing the metrics that the hybrid model improves upon. Overall, the hybrid model allows doctors and clinicians to get the most comprehensive result out of all three of our designed models in a drastically shorter amount of time than would be required for clinical diagnosis.

Table 4: Compares the results of the three proposed modules in this work: the social behavior based module, the facial feature based module, and the hybrid model.

| Statistic | Social Behavior Module | Facial Feature Module | Hybrid Model |
| --- | --- | --- | --- |
| Age Range | 2-8 | 2-8 | 2-8 |
| Accuracy | 96% | 92% | 87% |
| Sensitivity | 96% | 91% | 84% |
| Precision | 77% | 89% | 89% |
| Dataset Size | 1319 | 2836 | 125 videos, 235 images |
| $N_{ASD}/N_{non-ASD}$ | 1046/273 | 1418/1418 | 130/105 |

4. Conclusions

Three different modules for an ASD diagnosis based on a machine learning approach have been proposed in this research, and each module's performance has been compared to each other, with the hybrid model performing the best at an 87% accuracy. To the best of our knowledge, this is the highest reported accuracy based on our machine learning methods for this specific diagnosis. Although the single modules developed in this research are capable of providing a strong ASD diagnosis, our proposed hybrid model provides the most comprehensive and reliable diagnosis. The proposed architecture adds a unique applicability in a variety of scenarios. For example, the social behavior based module can be used when a patient only has ADOS data available and wants a quick diagnosis. However, certain areas of the world may not have as many ADOS materials or experts, so patients may use the facial feature based module. It is a simple procedure: record and send a video of the patient. Even though these scenarios may bring about a diagnosis, the most reliable and comprehensive diagnosis comes from utilizing both social behavior and facial feature data in our hybrid model. Not only does the hybrid model outperform the singular modules, but it is also more comprehensive, as it analyzes two areas of the body that ASD strongly affects.

We plan to continue enlarging our database by gathering more videos. We also plan to continue our research into clinical trials to identify areas of improvement and to continue development. Additionally, we propose for clinics to build up paired databases of ASD patients in the future. Since we only analyzed data from toddlers and children, we will expand our data to incorporate different age groups, like teenagers or adults, as ASD affects people of all ages. Finally, we plan to gather more data and build our architecture to analyze and diagnose different developmental disorders, like ADHD or Parkinson's disease.

**Acknowledgments**